\documentclass[doublecol]{epl2} 

\usepackage{amsmath}
\usepackage{amssymb}
\usepackage{epsfig}
\usepackage{graphics}
\usepackage{graphicx}
\usepackage{paralist}
\usepackage{hyperref}
\usepackage[all]{hypcap}
\usepackage{xcolor}

\usepackage{dcolumn}   
\usepackage{bm}        

\newcommand{\sub}{\raisebox{-3pt}}

\title{Micelle fragmentation and wetting in confined flow}
\shorttitle{Micelle fragmentation and wetting in confined flow} 

\author{Mona Habibi\inst{1} \and Colin Denniston\inst{1} \and Mikko Karttunen\inst{2}}
\shortauthor{M. Habibi \etal}

\institute{                    
  \inst{1}\!\!\!\!Department of Applied Mathematics, the University of Western Ontario, 1151 Richmond Street North, London, Ontario N6A\,3K7, Canada\,\,
  \inst{2}\!\!\!\!Department of Chemistry \& Waterloo Institute for Nanotechnology, University of Waterloo, 200 University Avenue West, Waterloo, Ontario, Canada N2L\,3G1
}

\abstract{We use coarse-grained molecular-dynamics (MD) simulations to investigate the structural 
and dynamical properties of micelles under non-equilibrium Poiseuille flow in a nano-confined geometry. 
The effects of flow, confinement, and the wetting properties of die-channel walls 
on spherical sodium dodecyl sulfate (SDS) micelles
are explored when the micelle is forced through a die-channel slightly smaller than its equilibrium size. 
Inside the channel, the micelle may fragment into smaller micelles. In 
addition to the flow rate, the wettability of the channel surfaces dictates whether the micelle 
fragments and determines the size of the daughter micelles: The overall behavior 
is determined by the subtle balance between hydrodynamic forces, micelle-wall interactions and self-assembly forces.}

\begin{document}

\maketitle

\section{Introduction}
  \begin{figure}     
            \onefigure[width=0.8\linewidth]{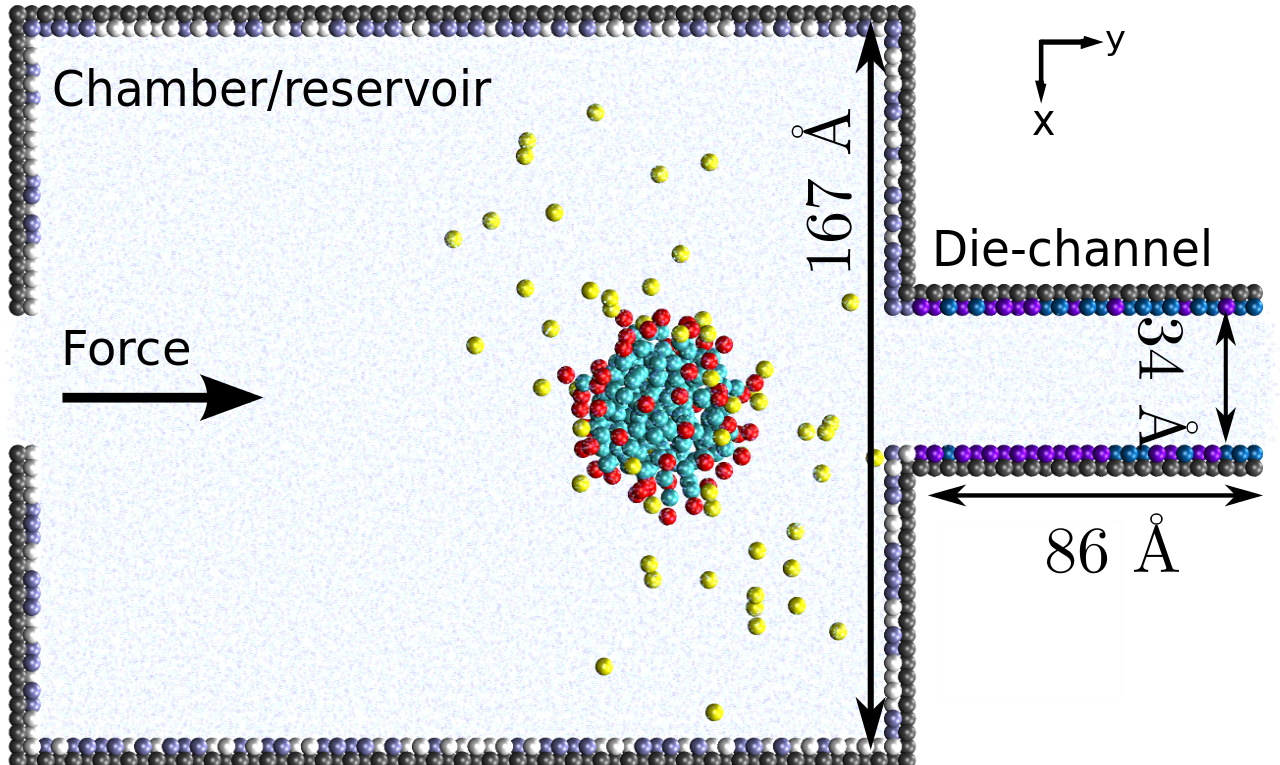} 
            \caption{(Color online) Cross-section in the $xy$-plane of the micelle in the 
die-channel geometry (initial configuration). Blue shadow points: water beads;  yellow spheres: counter-ions; 
            surfactants' heads: red; tails: cyan. 
In the chamber area, white and purple beads on the walls represent charged beads, 
with $\pm 0.2 \mathrm{e}$ charge, respectively. 
The beads on the inner wall layer in the narrow channel area are separately parameterized 
for non-, low-, and high-wetting surfaces to SDS micelles, Table~\ref{tab:wallwatere}. These beads are depicted in violet 
($+0.2 \mathrm{e}$) and blue ($-0.2 \mathrm{e}$). The system is periodic in the $y$- and $z$-directions. 
The inner size of the system is $167\times278\times185$ {\AA} in $x\times y \times z$. 
The inner width of the narrow channel is 34\,\AA, slightly less than the diameter of the equilibrated micelle (41\,\AA).
}
           \label{figchannel}
  \end{figure}
 The dynamics of deformable objects, such as micelles and vesicles, is fundamental in 
a wide range of applications from drug delivery to oil recovery~\cite{micelle-drugdelivery,sds-drugdelivery}. 
Typically, they are subjected to flow and extruded through micro-channels, thin capillaries, blood vessels 
and nanopores, where the sizes of self-assembled structures are similar to the distance between 
channel walls~\cite{vesicle-extrude-hope,vesicle-extrude-patty,vesicle-extrude-langmuir,polymer-stretchDPD-wettability}. 
As a result of surface absorption and flow, these objects break, alter their shapes 
and form new structures~\cite{vesicle-extrude-hope,vesicle-extrude-patty,vesicle-extrude-langmuir,vesicle-extrude-cg,polymer-stretchDPD-wettability,Adsorption-SDS-surface,Adsorption-SDS-surface2,Adsorption-CTAB-silica,surf-surface-Klein}. 
Due to the inherent non-equilibrium nature 
of these systems, 
theoretical description is very challenging 
as is tracking the behavior of an individual aggregate, or molecules forming them;  
soft self-assembled structures are a very active area of research\cite{vesicle-extrude-hope,vesicle-extrude-langmuir,pnas-wormmicelle,vesicle-extrude-cg,rbc4,Gompper-vesicle,santu2,wormicelle-die,polymer-stretchDPD-wettability,Adsorption-SDS-surface,Adsorption-SDS-surface2,Adsorption-CTAB-silica,surf-surface-Klein}.

     \begin{figure*}
                  \onefigure[width=0.95\linewidth]{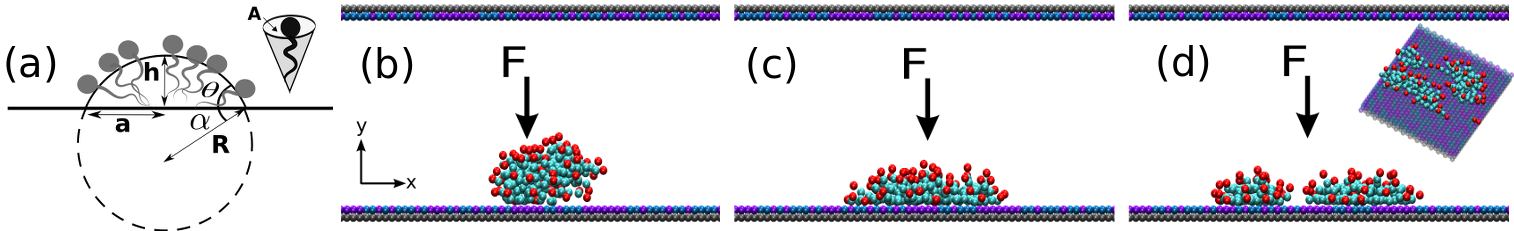} 
               \caption{(Color online) 
a) A bulb micelle on a surface. A surfactant occupies a cone shape volume with 
surface area $A$. Surface area of the micelle cap is then approximated by $S\!=\!\!gA$; $g$ 
is the number of SDS molecules forming the cap (Eq.~\ref{gcomp}). 
$R=20.5\pm0.3$ {\AA} is the radius of the micelle at equilibrium.
b-d) Response to an external force ($\mathbf{F}$) on a low-wetting surface. 
b) Initial configuration, (c) spreading after t=1.8\,ns, and d) fragmentation at t=10\,ns. 
The micelle breaks 
under the external force. 
After fragmentation, fragments are pushed away from each other. They distort and take a more stretched shape (inset in c)). This fragmentation is similar to the breaking up of micelles inside the channel when Poiseuille flow is applied (Fig.~\ref{figsnapshots}). The aggregate sizes can be estimated through surface tension measurements (Eq.~\ref{gcomp}). 
Counter-ions and water molecules are not shown.
}
              \label{figwetforce}
      \end{figure*}

Recent experiments show that extrusion can be used to size and reshape micelles in a very efficient way\cite{extrusion}. 
The physical mechanisms controlling the various stages of the process remain, however, unresolved.
 Computer simulations offer an alternative by providing a non-perturbative computational microscope to structure and dynamics. 
 They have become a key component in studies of rheology of micellar 
 solutions~\cite{shear-wormmicelle-pois-dpd}. Due to the time and length scales required, coarse-grained (CG) approaches \cite{cg1,martini-2007} 
 are needed; by reducing the degrees of freedom in the system, time and length scales can be expanded 
 to capture transport properties and shape transitions.  We performed coarse-grained 
 MD simulations to investigate the effects of Poiseuille flow, confinement, and the wetting properties of channel 
 walls on spherical  micelles when they are forced through a channel slightly smaller than their 
 equilibrium size (Fig.~\ref{figchannel}).  A sodium dodecyl sulfate (SDS) micelle was chosen as 
 the model because of its applications, e.g., in drug delivery systems \cite{sds-drugdelivery}. 
 We demonstrate that the interplay between hydrodynamics and wettability of narrow channel walls 
 determines whether micelles break up, and the sizes of resulting daughter micelles. 
Simulations were performed in two systems: 1) The {\it driven} system, the die-channel 
simulations under Poiseuille-like flow, where the fluid particles were driven by a constant-body 
force (Fig.~\ref{figchannel}); and  2) the {\it pushed} system, a slab-geometry simulation 
where we applied an external force on the SDS molecules to push the micelle toward the wall 
\textit{without} any flow (see Fig.~\ref{figwetforce}). In the driven simulation (die geometry) the velocity profile of the flow inside the channel was parabolic. Thus, as the micelle enters the narrow channel area, the micelle contacts the walls. In the second type of simulation, a perpendicular external force is applied only to the micelle, to replicate the situation in the die-channel where the micelle interacts with the surface. These two types of the simulations are not exactly the same, however the latter can be used to illustrate that the micelle fragmentation on surfaces is governed by the wall wetting.    

\section{Simulation details}

 The MARTINI force-field martini-v2.P~\cite{martini-2007,martini-polarwater} 
 was used for the interactions between the CG polarizable 
 water molecules, CG counter-ions, and CG SDS. Non-bonded interactions are described by the Lennard-Jones (LJ) 
 potential and Coulombic potential energy function~\cite{martini-2007,martini-polarwater}. 
 Bonds are described by a harmonic potential and a weak cosine-type harmonic potential 
 is used for the bond angles. 
In the MARTINI  force-field, on average 4 heavy atoms are represented by one interaction center (or "bead"). The LJ parameter is $\sigma=4.7$~{\AA} for all the MARTINI beads. A CG SDS has four beads; one charged bead represents the head group, and the tail group has three apolar beads. A positive charged bead represents an ion and a hydration shell around it. The polarizable MARTINI water \cite{martini-polarwater} has three linear beads 
 of equal masses (24 $\mathrm{\frac{g}{mol}}$) representing four water molecules. 
 Side beads have a partial charge of $q_\pm =\pm0.46\mathrm{ e}$ each ($\mathrm{WP,WM}$ respectively). The central bead ($\mathrm{W_{0}}$) is neutral.  Full details of the atomic structure of SDS, water, and counter-ion (Na$^{+}$) molecules, along with the full interaction parameters can be obtained from  \url{http://md.chem.rug.nl/cgmartini/} 
and as described in Refs.\cite{martini-2007,martini-polarwater}. 
  MARTINI 
has been used successfully to investigate the self-assembly of a large class of lipids and surfactants into vesicles and micelles \cite{martini-2007,martini-micelle,cgsdsmartini}. 

 
  Simulations were performed with the LAMMPS~\cite{lammps1} MD package. 
The velocity-Verlet algorithm with a 10 fs timestep was used to integrate Newton's equations. 
The neighbor list was updated every 2 steps with a cutoff distance of $15$~{\AA}. 
The dissipative particle dynamics (DPD) thermostat
\cite{DPD-mikko2003} 
was used to maintain 
an NVT ensemble at $300 \mathrm{K}$;  DPD conserves momentum and reproduces correct hydrodynamic 
behavior under flow conditions \cite{dpdsoddemann,cg1}.  The thermostat had a 
cut-off distance of 8~{\AA} and friction coefficient $3.62$ in reduced units 
($ 33.32 \mathrm{\frac{kg}{mol\, ns}}$) and was applied only to fluid particles in 
the direction perpendicular to the driving force ($x$ and $z$). 
Typical for thermostating a group of atoms in systems under a flow, 
the temperature was calculated after substracting out  spatially-averaged velocity 
field
\cite{thermoevans}.
Bond lengths between water beads were kept rigid with the SHAKE algorithm\cite{shake}.
The system is periodic in the $y$- and $z$-directions. 

 
 
   A snapshot of a micelle under Poiseuille flow is shown in Fig.~\ref{figchannel}. 
   Walls were constructed from two static layers (110 plane) of an fcc lattice with lattice 
constant 6.23\,\AA. The beads in the outer layer of walls are neutral, while the beads of 
the inner layer are randomly charged ($\pm0.2\mathrm{e}$) with total charge of zero. Without 
the wall charges, the polarizable MARTINI water freezes onto the surfaces
due to the alignment of dipoles 
(polarizable MARTINI water molecule is a polarizable dipole). This artificial freezing 
is eliminated by using a random arrangement of weakly charged wall beads. This provides 
realistic non-stick, non-slip surfaces for the polarizable MARTINI water model under flow. 
Table~\ref{tab:wallwatere} shows our new wall-water LJ interaction parameters. 
Charged water beads (WM, WP) have LJ interactions with inner charges wall beads with $\sigma=2.35$~{\AA}, 
while for all other interactions $\sigma=4.7$~{\AA}~\cite{martini-2007}. 
   
   \begin{table}[htb]
     \caption{ LJ interaction strengths ($\mathrm{\frac{kcal}{mol}}$) for wall-water and wall-surfactant.  
$\mathrm{Wall_{0}}$: neutral wall beads;  $\mathrm{Wall_{-}}$/$\mathrm{Wall_{+}}$: negatively/positively 
charged beads in the inner layer of the wall.}
      \label{tab:wallwatere}
      \begin{center}
     \begin{tabular}{l|ccr}
          Water: &$\mathrm{W_{0}}$	&$\mathrm{WM}$	&$\mathrm{WP}$\\\hline
         $\mathrm{Wall_{0}}$ & 0.04 & 0 &  0 \\
         $\mathrm{Wall_{-}}$ & 0.04 & 0.836 & 0.956  \\
        $\mathrm{Wall_{+}}$ & 0.04 & 0.956 & 0.836 \\\hline\hline
        	  Surfactant: &Head	&Tails	&$\mathrm{Na^+}$\\\hline
                  Non-wetting    & 0.05 & 0.02 & 0.05\\
                  Low-wetting    & 0.3 & 0.12 & 0.12 \\
                  High-wetting  & 0.5 & 0.2 & 0.5 \\
      \end{tabular}
     \end{center}
   \end{table}

\begin{figure}[htb]
          \onefigure[width=0.8\linewidth]{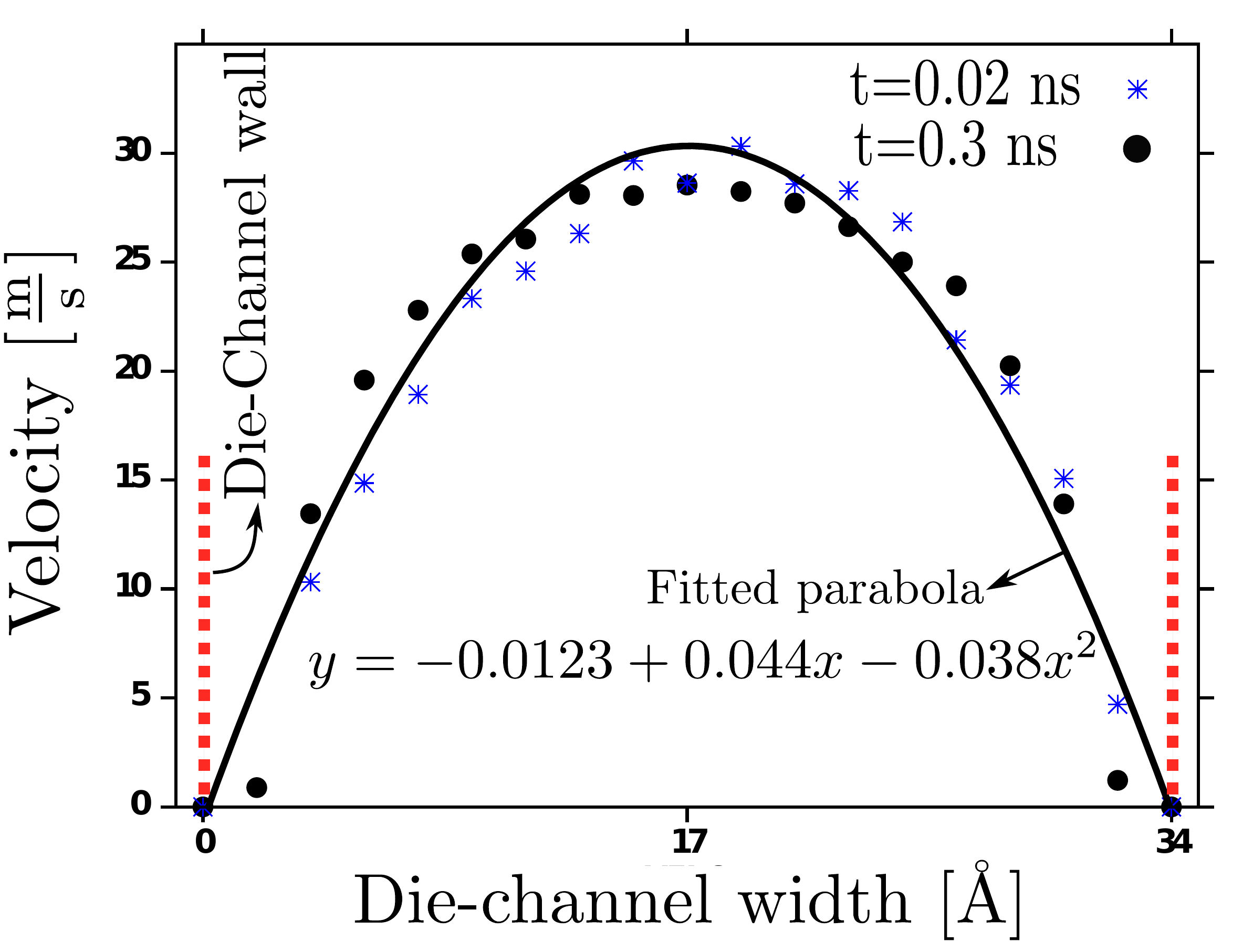}
         \caption{The velocity profile in the channel area 0.02\,ns (blue stars) and 0.3\,ns (black circles) 
after a body force of 0.2\,pN is turned on. The flow is a Poiseuille-like inside the channel. 
The velocity smoothly comes to zero at the walls.}
 	\label{profile}
 \end{figure}

\begin{figure}[htb]
          \onefigure[width=0.8\linewidth]{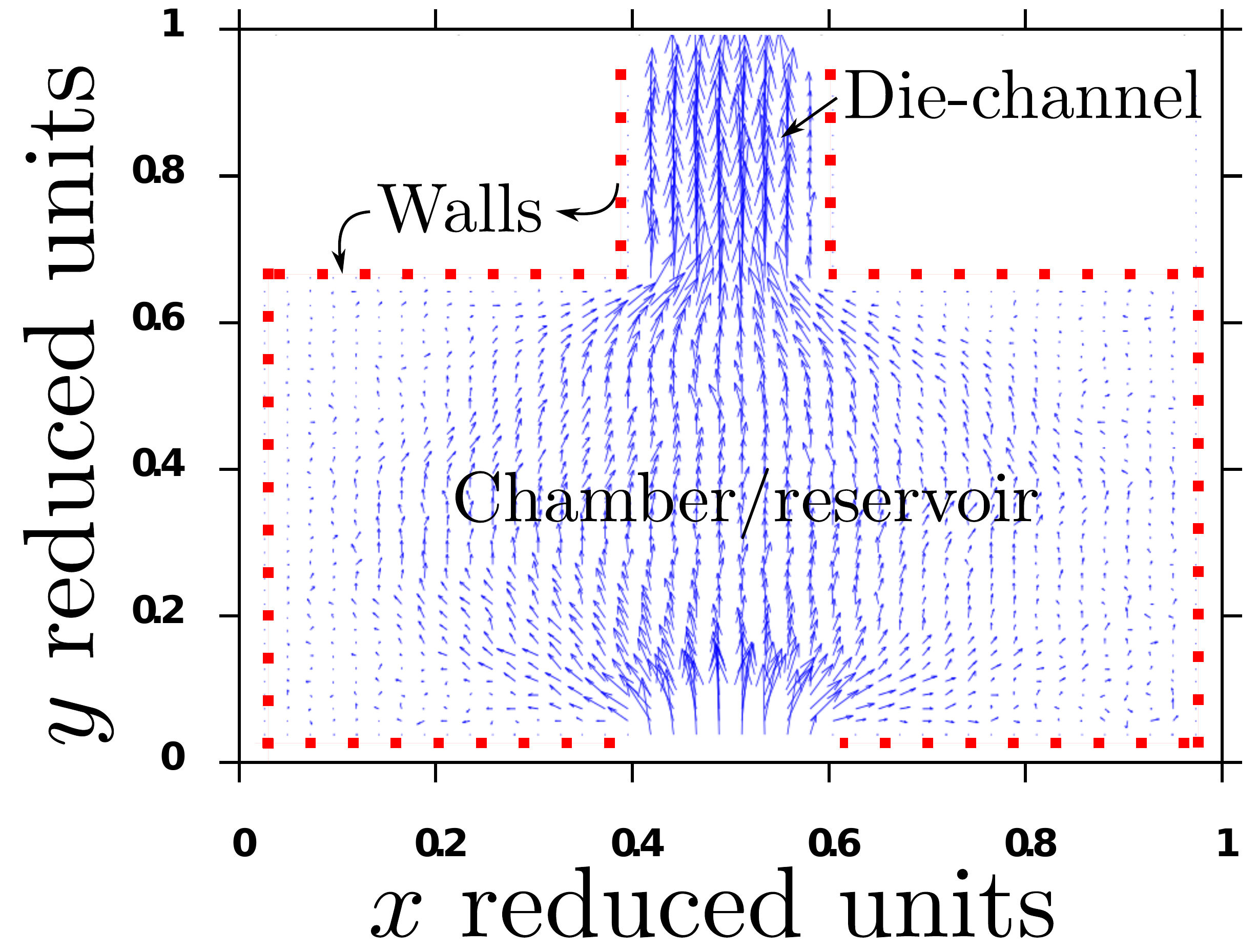}
         \caption{Vector plot of the flow in the system in the $xy$-plane.}
 	\label{vectorplot}
 \end{figure}

  The above interactions provide no-slip, no-stick boundary conditions
 as 
 demonstrated by the velocity profile of the flow inside the channel shown 
 in Fig.~\ref{profile} for a body force of $F=0.2$~pN. 
 Applying a 
 uniform-force on the system yields a Poiseuille-like flow.  
 The red dotted line shows the position of the channel walls and we see that the velocity 
smoothly comes to zero at the walls.  
 The fluid reaches the steady state in less than 0.3 ns. 
 Figure~\ref{vectorplot} shows a vector plot of the flow in the system in $xy$-plane 
in steady-state under a uniform force of 0.2~pN. 

 The equilibrium size of a spherical SDS micelle depends on the SDS molecular geometry and solution 
conditions such as temperature and counter-ions (Na$^+$ in this case).  
As such, it is a sensitive test of the force-field parameterization.  
The size was determined in separate simulations
to be 60 SDS molecules (the same number of counterions was present to ensure charge-neutrality). 
The spherical shape and size of this CG micelle are in good 
agreement with experiments~\cite{sdssize}, all-atom simulations \cite{Maria.SDS.2009} 
and a previous study on MARTINI SDS~\cite{cgsdsmartini}. This equilibrated spherical 
structure was used as the initial micelle in the simulations.

 Chamber length (see Fig.~\ref{figchannel} for the simulation geometry)
was varied from 100 to 192~{\AA}, the channel length from 86 to 236~{\AA}, 
 chamber width from 149 to 167~{\AA}, and the channel width from 24 to 36~{\AA}. 
 The applied force was varied over an order of magnitude, 0.034-0.2~pN. 
 The results in the rest of paper
 showcase the system with chamber length 192~{\AA}, channel length 86~{\AA}, and
 chamber width 167~{\AA}, but the same phenomena occurred independent of the parameters: The
 observations are general over a wide range of parameters, systems sizes and flow rates.

   To characterize equilibrium wetting properties of the walls, 
we  first studied the spreading of a spherical micelle droplet on the walls  
over various interaction strengths through measuring the surface tensions in 
a slab geometry. The micelle and counter-ions were solvated in a box of size 
$ 183.8\times 88.4\times 82 \, \mathrm{\AA^3} $. The density of water was 
$\rho=1060 \mathrm{\frac{kg}{m^3}}$. The micelle was centered at the top of 
the surface and slowly moved toward the surface until the surfactants' tails 
just touched the surface and then we let the system to equilibrate.

 The LJ interactions between walls and surfactants 
 were parameterized at three distinct levels corresponding to non-, low-, and 
 high-wetting surfaces, Table~\ref{tab:wallwatere}. The degree of wettability is 
 determined from contact angles of the micelle on the surface {\it at equilibrium}. 
Wetting is defined for the micelle as how well the micelle droplet spreads on a solid surface.
We would like to emphasize that in a non-wetting channel and in the absence of external forces, a micelle would never come into contact with the non-wetting surface. While for low- and high-wetting surfaces, depending on the wettability of the surface toward the micelle, the micelle droplet interacts with the surface. The balance of forces at the interface of the micelle and surface determines the macroscopic contact angle of the micelle. 
 The contact angle of a spherical micelle on an solid surface, $\theta_{s}$, is observed to obey Young's 
equation~\cite{wettingrev-degennes}
   \begin{equation}
     \cos{\theta_{s}}= \frac{\gamma_{\sub{\tiny{WALL,W}}}-\gamma_{\sub{\tiny{WALL,SDS}}}}{\gamma_{\sub{\tiny{W,SDS}}}},
     \label{thetas}
   \end{equation}
   where $\gamma_{\sub{\tiny{WALL,W}}}$, $\gamma_{\sub{\tiny{WALL,SDS}}}$, ${\gamma_{\sub{\tiny{W,SDS}}}}$ are the surface tensions of wall-water, wall-SDS, and water-SDS surface respectively. They were measured using the Kirkwood-Buff formula \cite{kirkwood-surfacetension}, $\gamma=\int\left[ P_{\perp}-P_{\parallel}\right]dr_\perp$ where $P_{\perp}=P_{yy}$ is the perpendicular pressure, $P_{\parallel}=\frac{1}{2}(P_{xx}+P_{zz})$ is the lateral pressure, and $r_\perp$ is the direction perpendicular to the interface. To measure ${\gamma_{\sub{\tiny{W,SDS}}}}$, the full 3D pressure field was obtained for the spherical interface~\cite{surfacetension-MD3d}. For various set of LJ interaction parameters, the contact angle was obtained using Eq.~\ref{thetas} 
and agrees with direct observation. 
Note that the contact angles are measured for the micelle, not for individual SDS molecules.
 As the simulation proceeds, the micelle interacts with the wall and depending on the interactions 
  between the wall and surfactant it either wets the surface to form a cap, forms a bulb shape micelle 
  or totally detaches from the surface. These correspond to high-, low-, 
  and non-wetting surfaces.
The water(neutral bead)/wall(neutral bead) LJ interaction energy was chosen as our reference point, 
and the SDS/$W$ and Na$^{+}$/$W$ LJ interactions were properly scaled with respect to $W$/wall 
energy interactions which gives the non-wetting wall. By increasing the LJ interactions by a factor of 10, the high-wetting surface were formed. 
The LJ parameters for
are shown in Table~\ref{tab:wallwatere}.

  Next, flow was applied to study the behavior of micelles using the above three different channel surfaces. 
After equilibration,
a uniform body-force of 0.2\,pN in the $y$-direction was 
applied to each fluid particle, including the micelle, ions and water beads. 
The force pushes the particles and yields a Poiseuille-like flow in the channel.

\section{Results}
   \begin{figure*}[htb]
          \onefigure[width=0.85\linewidth]{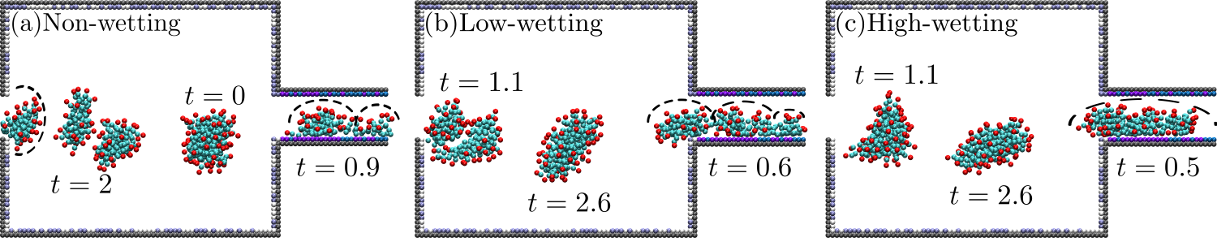}
         \caption{(Color online) Snapshots of the system with different surfaces under the body-force of 0.2 pN. Micelles are shown during the first time passage at different times in the simulation. Times are in ns after the driving force is turned on. For clarity, water molecules and ions are not shown. From left to right: Non-wetting, low-wetting, and high-wetting channels. Micelles adopt a cigar shape conformation to pass through the channel. Inside the channel, the SDS molecules contact the surface. The micelles break up on the non/low-wetting surfaces, and may re-assemble in the chamber area. Inside the high-wetting channel, the micelles spread on the wall and crawl as one whole structure. Inside the channel, a dashed line is used to show a micelle fragment. In a) and b) multiple fragments are formed, while in c) the micelles does not break. }
 	\label{figsnapshots}
 \end{figure*}
  Snapshots of the simulations with different channel wettabilities are shown in Fig.~\ref{figsnapshots}. The micelle is shown at three different times during its first passage through the channel. Figure~\ref{figsnapshots} illustrates how the micelle shape and size fluctuate under flow. Depending on the flow rates and the wetting properties of the walls, the micelle behaves differently. The micelle fragments on non/low-wetting surfaces (see Fig.~\ref{figsnapshots}). On a high-wetting surface, under the same flow conditions, the micelle passes through the channel as one whole structure.
 When in the chamber before the first passage, the flow velocity determines a micelle's 
 shape and radius independent of the surface properties of the walls. 
 Once pushed by the flow field inside the channel, 
 the wetting properties of the channel surfaces have a strong influence on its dynamics: 
 With highly wetting surfaces, the micelle totally spreads on the surface. 
 The surfactants wet one of the walls (simultaneous fragmentation over high-wetting surfaces was not
 observed) and the micelle crawls through the channel. It finally exits and 
 re-enters the chamber as one structure (Figs.~\ref{figchannel} \& \ref{figsnapshots}). Inside the chamber, the micelle regains its spherical shape. On the other hand, if the walls are low-wetting/non-wetting, micelle fragmentation occurs. Inside the non-wetting/low-wetting channel, the micelle splits into daughter micelles with varying number of surfactants. 
 Given sufficient time, when the smaller fragments re-enter the chamber, some or all may combine and re-assemble into one or more micelles with different sizes. Micelle fragments follow the path through the narrow channel almost with the same pattern, elongating along the direction of the applied force to pass through the channel.

Since a flowing micelle can break into multiple fragments, with each having a different 
number of SDS molecules, we need to identify which SDS belongs to which daughter 
micelle in order to quantify the effect. 
Similar to the method presented in Ref.~\cite{Maria.SDS.2007}, 
for each pair of SDS molecules we measure the distances between each of the corresponding 
beads on each of the two molecules, and the distance between the molecular centers of mass.
Any two SDS  molecules are classified to belong to the same micelle if they meet one of these criteria:
   \begin{inparaenum}[(i)]
          \item one of the calculated distances is less than $R_{cut_1}=6$ \AA,
          \item any of two distances are shorter than $R_{cut_2}=9$ \AA , and
          \item any of three distances are shorter than $R_{cut_3}=12$ \AA 
   \end{inparaenum}.
 The cutoff distances were chosen after inspecting the configurations of micelles. These criteria give results consistent with the number of micelles that were seen in visualizations. We ignored micelles with less than $5$ SDS.

\begin{figure}[htb]
                    \onefigure[width=0.75\linewidth]{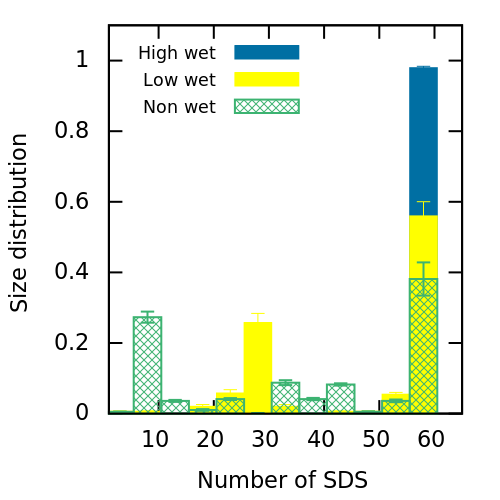}  
                    \caption{(Color online) Micelle size distribution inside the channel for 
                    the first passage for non- (criss-cross green), low- (solid yellow), and high- (solid blue) wetting channel surfaces. The significant peak for low-wetting surfaces shows that the daughter micelle size peaks at 30. 
                    For non-wetting surfaces it is more probable to have daughter micelles of size 10. Decreasing wettability leads to more micelles with smaller sizes. The error bars  are the standard deviations for the numbers of micelles in each bin.}
                   \label{figsizedist1}
              \end{figure}

    The micelle fragment size (i.e. number of SDS molecules) distribution after the first passage through the channel is shown in Fig.~\ref{figsizedist1}. The distribution was obtained over 10 simulations. Since the initial size of the micelle was 60 and the micelles spend some time inside the channel before breaking up, the largest peak is 60 for all cases. For the high-wetting surface, micelles of size 60 pass through the channel without any noticeable fragmentation. Figure~\ref{figsizedist1} shows that as the wettability of the surface decreases, it is less probable to have micelles of size 60. For the low-wetting surface, the micelle splits into daughter micelles inside the channel with an average size between 25 and 35 (second significant peak). For the non-wetting surface, smaller micelles with an average size of 10-15 are formed; if the micelle repeatedly passes through the non-wetting channel, one expects to see multiple micelles with an average size of 10. For low-wetting surfaces, less micelles with larger sizes will be formed. Thus, if the micelles are pushed through long channels, they are likely to emerge at the end in a micellar solution which will consist of a relatively mono-disperse size distribution of micelles.     
         
    Over various sizes of the die simulations, the fragmentation occurs when the micelle is inside the channel area and interacts with the channel surfaces. Thus, the micelle fragmentation is not only due to the hydrodynamic forces. In fact, the wetting properties of the surfaces play a major role in size and morphology of micelles on surfaces, even at equilibrium. Adsorption and the aggregation of SDS molecules into hemimicelles(cap-shape micelle) on various substrates have been reported repeatedly in experiment and simulations~\cite{Adsorption-SDS-surface, Adsorption-SDS-surface3}.

    The fragmentation of micelles into smaller bulb shaped micelles under flow is 
a result of an interplay between hydrodynamic forces on the micelle, the wall-micelle interactions, 
and self-assembly forces. 
    As the micelle enters the channel, it is pushed against the surface of the channel. 
It is possible to predict the sizes of the resulting micelle fragments if the contact angle is known. 
To test this, we revisit the slab geometry to classify surface-micelle interactions in terms of classical wetting properties in the non-equilibrium forced condition. This setup is used to measure the surface tensions at the micelle-surface interface when the micelle is pushed against the surface. This is similar to the case where the micelle flows inside the channel and the lateral hydrodynamic forces push it against the walls.

    We measure $ \gamma_{\sub{\tiny{WALL,W}}}$ and $\gamma_{\sub{\tiny{WALL,SDS}}}$, and then calculate the optimum contact angles using Young's equation, Eq.~(\ref{thetas}).  One should notice that in equilibrium, when the micelle touches the wall the interaction is primarily with the head groups, i.e., they dominate the equilibrium wetting properties. However, when the micelle is forced against the wall, the hydrocarbon groups of the tails start to interact with the wall atoms. The wetting properties of the heads and tails are naturally reversed due to the hydrophobic and hydrophilic properties of tails and heads. As a result, the wetting properties of the tails dominate fragmentation.  For a spherical micelle, the surface area is $S=4\pi R^2=gA$, where $g$ is the number of surfactants in a micelle. $A=\frac{4 \pi R^2}{N_{sds}}$ is the surface area a surfactant occupies in a spherical micelle consisting of $N_{sds}$ surfactants ($N_{sds}=60$) and $R$ is the radius of the micelle in equilibrium. When the micelle is pushed against the surface, it takes the shape of a spherical cap. We can then approximate the surface area of the micelle as the surface area of the cap, $S=2\pi Rh$, where $h$ is the height of the cap. By knowing the contact angle $\theta$, $h$ can be defined from $\alpha=\pi/2-\theta=\arcsin(\frac{R-h}{R})=\arccos(\frac{a}{R})$ 
(see Fig.~\ref{figwetforce}a). 
Finally, we estimate the number of SDS molecules ($g$) in a micelle fragment from
   \begin{equation}     
       g=\frac{2 \pi R h}{A}=\frac{2 \pi R^2}{A}\left[1-\left(\frac{\gamma_{\sub{\tiny{WALL,W}}}-\gamma_{\sub{\tiny{WALL,SDS}}}}{\gamma_{\sub{\tiny{W,SDS}}}}\right)\right].
       \label{gcomp}
   \end{equation}

%


  Using the slab geometry (Fig.~\ref{figwetforce}b-d), we applied a constant force evenly only on the SDS beads to push the surfactants toward the solid surface. The applied force was similar to the average hydrodynamic force exerted on each channel wall molecule in the die simulation. The sizes of the micelle fragments were computed for each case. Data were averaged, starting from the time that the micelle totally spreads over the surface until it breaks for the first time. During this time, the micelle has a more cap shape structure on the surface(see Fig.~\ref{figwetforce}~b,c)). After the micelle breaks up, the fragments are pushed away from each other and deformed. Eq.~(\ref{thetas}) was used to obtain $\theta$. Figures~\ref{figwetforce}b-d show the time evolution of the micelle on the low-wetting surface under external force $F=5.52 \mathrm{pN}$ for $10$ ns. We measured $\gamma_{\sub{\tiny{WALL,SDS}}}= 54.6\pm6.0, 49.2\pm6.2, 41.4\pm 7.4 \mathrm{\frac{mN}{m}}$ for the non-, low-, and high-wetting cases, respectively. Our calculations show that the optimum number of SDS in the caps are about $g_{c}=7.8\pm18.9,24.6\pm19.4,49.3\pm23.1$, respectively. These numbers are in good agreement with the size distribution of micelles in the die simulation (Fig.~\ref{figsizedist1}). 
    
Our results illustrate that the fragment size distribution is controlled by the wall wetting 
when the micelle interacts with the surface. In equilibrium, a micelle will never 
spontaneously approach a non-wetting wall, unless it is being pushed onto the wall. 
When the channel is wider than the diameter of the micelle, 
it can pass through without interacting with the walls. 
Then, the wetting properties of the wall do not 
play a major role in micelle dynamics.


This simple model explains that the fragmentation of the micelle is the result of  
surface tension of the interfaces, and self-assembly and hydrodynamic forces. 
Thus, by controlling the wettability of the surface, size distribution can be controlled. 
Note that even in equilibrium, if a micelle wets a surface, fragmentation occurs due to the 
balance between the surface tensions at the interface of micelle/water and wall, and the forces within SDS molecules.
   
\section{Summary}
 To summarize, CG MD simulations of SDS micelles in the die-extruder geometry were performed. 
 The behavior of a micellar solution under Poiseuille-like flow in the presence of differently wetting surfaces was investigated. 
 Independent of the wetting properties of the channel walls, micelles were always pushed toward them 
 (the micelle was slightly larger than the channel width, Fig.~\ref{figchannel}). Micelle fragmentation inside the channel 
 is controlled by the wettability of the channel walls: Fragmentation can be induced by decreasing wall wettability. 
 For high-wetting surfaces with the same applied flow velocity, the tails of the SDS molecules spread on one of the walls and the 
 micelle crawls along the channel without splitting. After leaving the channel, 
 the micelle fragments may recombine into one whole micelle upon entering the chamber area.  The hydrodynamic forces need to be large enough to push the micelle into the channel, while small enough that the micelle has sufficient time in the channel to interact with the walls.  Once these constraints are satisfied, the details of the flow velocity or pressure drop are not, in fact, that important.
  The number of SDS molecules in the micelle fragments, 
 approximated from the macroscopic contact angle, was in good agreement with the size distribution of 
micelle fragments of the die simulations. Our analysis and simulations show that micelle fragmentation 
and the size of the daughter micelles, can be controlled by varying the wetting properties of the surfaces. 







\acknowledgments

We thank the Natural Science \& Engineering Research Council of Canada for financial support and 
Compute/Calcul Canada for computing facilities.


\end{document}